\documentclass[12pt]{article}
\usepackage{epsfig,amssymb,amsmath,psfrag,epstopdf,color}
\usepackage{breqn}

\allowdisplaybreaks[4]
\numberwithin{equation}{section}

\def\beq{\begin{equation}}
\def\eeq{\end{equation}}
\def\bsp#1\esp{\begin{split}#1\end{split}}
\newcommand{\be}{\begin{equation}}
\newcommand{\ee}{\end{equation}}
\newcommand{\bea}{\begin{eqnarray}}
\newcommand{\eea}{\end{eqnarray}}

\def\to{\rightarrow}

\newbox\charbox
\newbox\slabox
\newbox\ourfigbox
\def\s#1{{      
        \setbox\charbox=\hbox{$#1$}
        \setbox\slabox=\hbox{$/$}
        \dimen\charbox=\ht\slabox
        \advance\dimen\charbox by -\dp\slabox
        \advance\dimen\charbox by -\ht\charbox
        \advance\dimen\charbox by \dp\charbox
        \divide\dimen\charbox by 2
        \raise-\dimen\charbox\hbox to \wd\charbox{\hss/\hss}
        \llap{$#1$}
}}

\def\ksl{\not{\hbox{\kern-2.3pt $k$}}}

\def\e{\epsilon}

\def\Ord{{\cal O}}

\def\spa#1.#2{\left\langle#1\,#2\right\rangle}
\def\spb#1.#2{\left[#1\,#2\right]}
\def\lor#1.#2{\left(#1\,#2\right)}
\def\sand#1.#2.#3{%
\left\langle\smash{#1}{\vphantom1}^{-}\right|{#2}%
\left|\smash{#3}{\vphantom1}^{-}\right\rangle}

\def\n{n}    

\def\del{\partial}

\def\n3lo{N$^3$LO}

\newfont{\scyr}{wncyr10 scaled 550}

\newcommand{\dd}{\mathrm{d}}
\newcommand{\nn}{\nonumber}
\newcommand{\brk}{\nonumber\\&}

\newcommand{\mcdot}{\!\cdot\!}
\newcommand{\mi}{\mathcal{I}}
\newcommand{\mj}{\mathcal{J}}

\newcommand{\topo}{\mathrm{Topo}}

\begin{document}

\catcode`\@=11
\font\manfnt=manfnt
\def\Watchout{\@ifnextchar [{\W@tchout}{\W@tchout[1]}}
\def\W@tchout[#1]{{\manfnt\@tempcnta#1\relax%
  \@whilenum\@tempcnta>\z@\do{%
    \char"7F\hskip 0.3em\advance\@tempcnta\m@ne}}}
\let\foo\W@tchout
\def\dubious{\@ifnextchar[{\@dubious}{\@dubious[1]}}
\let\enddubious\endlist
\def\@dubious[#1]{%
  \setbox\@tempboxa\hbox{\@W@tchout#1}
  \@tempdima\wd\@tempboxa
  \list{}{\leftmargin\@tempdima}\item[\hbox to 0pt{\hss\@W@tchout#1}]}
\def\@W@tchout#1{\W@tchout[#1]}
\catcode`\@=12


\thispagestyle{empty}

\null\vskip-60pt \hfill
\begin{minipage}[t]{42mm}
SLAC-PUB-16183\\
\end{minipage}
\vspace{5mm}

\begingroup\centering
{\Large\bfseries\mathversion{bold}
On the calculation of soft phase space integral}%
\vspace{7mm}

\begingroup\scshape
Hua~Xing~Zhu\\
\endgroup
\vspace{5mm}
\begingroup\small
\emph{SLAC National Accelerator Laboratory, Stanford University, Stanford, California, 94025, USA}
\endgroup

\vspace{0.4cm}
\begingroup\small
E-mail: {\tt hxzhu@slac.stanford.edu}.
\endgroup
\vspace{0.7cm}

\textbf{Abstract}\vspace{5mm}\par
\begin{minipage}{14.7cm}
The recent discovery of the Higgs boson at the LHC attracts much
attention to the precise calculation of its production cross section
in quantum chromodynamics.  In this work, we discuss the calculation
of soft triple-emission phase space integral, which is an essential
ingredient in the recently calculated soft-virtual corrections to Higgs
boson production at next-to-next-to-next-to-leading order. The main
techniques used this calculation are method of differential equation
for Feynman integral, and integration of harmonic polylogarithms.
\end{minipage}\par
\endgroup

\newpage



\section{Introduction}
\label{IntroSection}

To maximize the physics outcome of the LHC program, theoretical
predictions for signal and background processes have to match the
unprecedented accuracy of experimental measurements. For
QCD initiated processes, a standard approach to improve the accuracy of
theoretical prediction is by calculating perturbative radiative
corrections beyond Leading Order~(LO)
approximation. At Next-to-Leading Order~(NLO), such an calculation
consists of virtual correction, which involves virtual parton
circulated in the loop, and real correction, which involves real on-shell parton emission. When considering corrections beyond NLO, mixed real-virtual correction is also required.
 In the past few decades, significant efforts have been made to
 develop  analytic techniques for loop amplitude calculation. In
 particular, a variety of methods have 
been developed for the analytic calculation of loop integral,
cf.~\cite{Smirnov:2012gma} and references therein. For phase space
integral, techniques for analytic calculation are considerably less
developed. An obvious reason is that experimental measurements often
involve complicated phase space cuts, which are very difficult to
implement analytically. The standard approach is to perform the phase
space integral numerically, keeping in mind that appropriate
subtraction terms have to be added in order to render the integral
finite~\cite{hep-ph/9512328,hep-ph/9605323}. In that case analytic
inclusive phase space integrals can be used in the construction of 
infrared subtraction terms for Next-to-Next-to-Leading Order~(NNLO)
calculation~\cite{hep-ph/0311276}. Moreover, 
when appropriate subtraction method was not yet available, analytic
integral was the only method for obtaining cross section or
distribution at NNLO at hadron
collider~\cite{NUPHA.B359.343,hep-ph/0201206,hep-ph/0207004,hep-ph/0302135,hep-ph/0306192}. 

Recently, an important step towards the calculation of Higgs
production at Next-to-Next-to-Next-to-Leading Order~(\n3lo) has been taken in
Refs.~\cite{1403.4616,1412.2771}, in which the soft-virtual
corrections are obtained. The soft-virtual corrections are cross
section in the threshold limit, where radiation in the final state
is constrained to have low energy. Since the radiation energy is
low, calculation can be done in the eikonal limit, where  QCD partons
are emitted from light-like Wilson lines which parameterize the
directions of incoming partons. Calculation of soft-virtual
corrections at \n3lo requires the calculation of pure virtual
corrections through to three loops, and soft phase space integrals
with one, two, and three partons in the final state. For soft phase
space integrals, the ones with three partons in the final state are
the most challenging one. Such triple soft phase space integrals are
first calculated in the pioneering work~\cite{1302.4379}, using
various powerful techiques for Feynman integrals, ranging from
Melin-Barnes transformation to symbols and coproduct. 

In Ref.~\cite{1412.2771}, a completely different approach is
taken. Auxiliary integrals are introduced in the intermediate steps, as
well as nontrivial dependence on \emph{internal} auxiliary scales. The
auxiliary integrals are then solved by the method of differential
equation for Feynman integral. Finally the auxiliary scales  are integrated over and the
desired  triple soft phase space integrals are obtained. This approach
makes heavy use of the property  of scale invariance for eikonal
integral. It is worth mentioning that the triple soft phase space
integrals are single scale integrals with trivial dependence on the
single scale, therefore prohibit direct application of the method of
differential equation. By introducing the auxiliary integrals,
additional nontrivial scales are introduced into the problem, and
nontrivial system of differential equation can be derived. While
differential equation approach to single scale integrals have been employed
before in different context~\cite{1312.2588,1408.5134} by introducing
\emph{external} scales, we find that auxiliary scales being
introduced \emph{internally} in the current calculation is
particularly interesting, and significantly simplifies the calculation
at all stages. In this paper we document the detail of such a calculation,
in the  hope that it will be useful in other problems.

This paper is organized as follows. In Sec.~\ref{sec:diff-equat-appr},
we explain the method in detail, using the NNLO double soft emission
phase space integral as example. In Sec.~\ref{sec:triple}, we apply
this method to triple soft phase space integrals. In
Sec.~\ref{sec:conclusion-outlook} we give a brief conclusion and
outlook. Detail for solving the system of differential equation for
triple soft integrals is given in Appendix.~\ref{sec:solv-diffr-equat}.

\section{Differential equation approach to single scale soft phase space integral}
\label{sec:diff-equat-appr}

\subsection{Introducing the method}

The method of differential equation~\cite{PHLTA.B259.314, PHLTA.B267.123,hep-ph/9912329} is powerful  in the calculation of
Feynman integrals with multiple scales. For single scale problem, the
dependence on the scale is trivial from dimensional analysis. The
nontrivial part of the integral is a function of dimensional
regularization parameter only~\footnote{We work in dimensional
  regularization with space-time dimension continued to $d$ dimension
  throughout this work.}. Thus the method of differential equation can
not be applied directly. In some cases, one can introduce external
auxiliary scales such that nontrivial differential equation for
Feynman integral can be derived. The integrals with additional
external auxiliary scales can then be related to the original integrals
by taking limit of the auxiliary external
scales~\cite{1312.2588,1408.5134}.  Unfortunately, the calculation of
Feynman  integral becomes more complicated as the number of scales increased. 

In this work, we solve single scale soft phase space integral by
introducing \emph{internal} auxiliary scales. Differential equations
with respect to these internal scales are derived and solved. Then the
internal auxiliary scales are integrated over, and the desired single
scale integral is recovered. There are several advantages in this
approach. First, the internal auxiliary scales will be introduced by
inserting a delta function. Some of the momentum integral can be
integrated out trivially by the delta function. Effectively, the number of loops involved
in the calculation is reduced by one. Reducing the number of loop significantly simplifies the complexity of Intergration-By-Parts~(IBPs)
reduction~\cite{PHLTA.B100.65,NUPHA.B192.159,hep-ph/0102033}. Second,
the property of scale invariance for eikonal integral can be fully
exploited after the introduction of internal auxiliary scales. The
consequence is that auxiliary integral only depends on a scale
invariant combination of the auxiliary scales. Therefore the
functional dependence of auxiliary integral on the auxiliary scales
has very restricted form.  Third, the internal
auxiliary scales are introduced in a way that integrating over them
is also trivial. In fact, only simple interative integrals of
harmonic polylogarithms~(HPLs)~\cite{hep-ph/9905237} are  needed for
this calculation, in contrast to the extensive uses of multiple
polylogarithms in Ref.~\cite{1302.4379}.

The single scale soft phase space integrals discussed in this work are
related to the soft-virtual corrections for color singlet particle production
at hadron collider. Perhaps the most important example is the
soft-virtual corrections for Higgs production~\cite{1403.4616,1412.2771}. Consider the process
\begin{align}
N_1 (P_1) + N_2(P_2) \to H (q) + X \,.
\end{align}
Thanks to QCD factorization, the hadronic cross section for this process
can be written schematically as a convolution of parton distribution
functions and partonic cross section
\begin{align}
  \sigma = f \otimes f \otimes  \hat{\sigma} \, .
\end{align}
The partonic cross section $\hat{\sigma} (z)$ depends on
 partonic threshold variable $z = M^2_H/\hat{s}$~\footnote{It also
   depends on Higgs mass $M_H$, top-quark mass $M_t$, renormalization scale
   $\mu_R$ and factorization scale $\mu_F$, which are irrelevant
   to our discussion.}, where $M_H$ is the mass of Higgs boson, and
 $\sqrt{\hat{s}}$ is partonic center-of-mass energy. The soft-virtual
 corrections are referred to the leading term of $\hat{\sigma}(z)$ in the expansion of $z$
 around $1$. Since the energy of final state radiation is $E_X \sim M_H
 ( 1 - z)/2 + \Ord \big( (1-z)^2 \big) $, calculation of soft-virtual
 corrections is equivalent to calculation of differential distribution
 in $E_X$, in the limit where $E_X$ is
 much smaller than all other scales in the problem. In this limit,
 the leading power contributions in $(1-z)$ come from the partonic
 process
 \begin{align}
   g(p_1) +  g(p_2) \to H(q) + X \, .
 \end{align}
In QCD perturbation theory, $X$ contains various combination of
massless QCD parton. For example, at NNLO, the double-emission
contribution contains the process
\begin{align}
  g(p_1) +  g(p_2) \to H(q) + g(k_1) + g(k_2) \, ,
\end{align}
while the triple-emission contribution at \n3lo contains the process
\begin{align}
    g(p_1) + g(p_2) \to H(q) + g(k_1) + g(k_2) + g(k_3) \,,
\end{align}
where $p^2_i = k^2_i = 0$.
The phase space integrals in the limit $E_X \ll M_H$ are single
scale integrals, whose dependence on $E_X$ can be determined uniquely
by dimensional analysis. The main purpose of this work is the
calculation of these integrals.

We describe in detail the method mentioned above, taking
double-emission soft phase space integrals as an example. After
application of IBPs reduction~\footnote{We use both \texttt{REDUZE 2}~\cite{1201.4330}
  and \texttt{LiteRed}~\cite{1212.2685} for IBP
  reduction.}, the result for double-emission contribution to NNLO
soft-virtual corrections can be expressed in terms of two nonzero
two-loop master integrals,
\begin{align}
  I_1 = & e^{2\e \gamma_E} \int\! \frac{\dd^d k_1 \, \dd^d k_2 }{ \pi^{d-2} } \,\delta_+ ( k_1^2) \,\delta_+ (k_2^2)
  \,\delta \big( 2 E_X M_H - (p_1 + p_2) \mcdot
  (k_1 + k_2) \big) \, ,
\nn \\
  I_2 = & e^{2\e \gamma_E} \int\! \frac{\dd^d k_1 \, \dd^d k_2 }{\pi^{d-2} } \, \frac{\delta_+ ( k_1^2) \,\delta_+
    (k_2^2) \,\delta \big( 2 E_X  M_H - (p_1 + p_2) \mcdot
  (k_1 + k_2) \big)}{ (k_1 + k_2)^2\, (2\, p_1 \mcdot k_1) \, (2\, p_2
  \mcdot k_2)} \, ,
\label{eq:2}
\end{align}
where $\gamma_E = 0.577216\ldots$.
The integrals are defined in $d = 4 - 2\,\e$ dimension to regulate infrared
singularities. The dependences of $I_1$ and $I_2$ on the kinematical
variables $E_X$, $M_H$, $p_1$ and $p_2$ are highly constrained. From
dimensional analysis, we have
\begin{align}
  I_1 \propto (E_X \, M_H)^{d - 3} \,,\qquad   I_2 \propto
  \frac{1}{p_1\mcdot p_2}(E_X \, M_H)^{d-5} \,,
\end{align}
where we have also made use of the fact that the scale
transformation
\begin{align}
p_1 \to \lambda_1 \, p_1 \, , \qquad p_2 \to \lambda_2 \, p_2 
\label{eq:3}
\end{align}
for the soft integrals is not anomalous, {\it i.e.}, anomalous term of
the form $(p_1 \mcdot p_2)^{a \e}$ is always absent. In the rest of this
paper we will set $2 E_X  M_H = 1$ for simplicity because their
dependence are known exactly. We will also let $p_1 \mcdot p_2 = 2$,
and introduce the standard notation
\begin{align}
  p_1 \mcdot k = k^+ \,, \qquad p_2 \mcdot k = k^- \,.
\end{align}
As explained above, $I_1$ and $I_2$ are
single scale integrals. Therefore the method of differential equation
for Feynman integral can not be applied directly. The solution  is to
introduce \emph{internal} scales by inserting an unit operator
\begin{align}
  1 = \int\! \dd^d l \, \delta^{(d)} \big( l - (k_1 + k_2) \big)
\end{align}
into the original integrals,
\begin{align}
  I_1 = & \, e^{\e \gamma_E} 
  \int\! \frac{\dd^d l}{\pi^{d/2-1}} \,\Theta(l^2)\,  \delta \big( 1 - (p_1 + p_2) \mcdot l
  \big) \, \mi_1 ( l, p_1, p_2) \, ,
\nn \\
  I_2 = &  \, e^{\e \gamma_E} 
  \int\! \frac{\dd^d l}{\pi^{d/2-1}} \,\Theta(l^2)\, \delta \big( 1 - (p_1 + p_2) \mcdot l
  \big) \, \frac{ \mi_2 ( l, p_1, p_2) }{l^2} \, ,
\label{eq:13}
\end{align}
where we have introduced the auxiliary integrals
\begin{align}
  \mi_1 ( l, p_1 ,p_2) = &\, e^{\e \gamma_E}\int\! \frac{\dd^d k_1}{\pi^{d/2-1} }\,
  \delta_+(k^2_1) \, \delta_+\big( (l - k_1)^2 \big) \,,
\nn
\\
  \mi_2 ( l, p_1 ,p_2) = & \,e^{\e \gamma_E} \int\! \frac{\dd^d k_1}{\pi^{d/2-1} }\,
   \frac{\delta_+(k^2_1) \, \delta_+\big( (l - k_1)^2 \big)}{(2 \, p_1
    \mcdot k_1 )\, \big(2 \, p_2 \mcdot (l-k_1)\big)} \,.
\label{eq:1}
\end{align}
Integrals of the form in Eq.~(\ref{eq:1}) are first introduced in
the context of fully exclusive soft function in ref.~\cite{1105.5171}, and
have been used in the calculation of threshold soft function for direct
photon production~\cite{1201.5572} and boosted top-quark pair production~\cite{1207.4798}. We
will see later that they are particularly convenient for evaluating
soft phase space integral. 

An
important feature of the auxiliary integrals in Eq.~(\ref{eq:1}) is
that they transform in a simple manner under Eq.~(\ref{eq:3}). The
scale symmetry and dimensional analysis determine that
\begin{align}
  \mi_1 (l,p_1,p_2) \propto & (l^2)^{(d-4)/2} f_1\left( y \right)\,,
\nn \\
  \mi_2 (l,p_1,p_2) \propto& \frac{(l^2)^{(d-4)/2}}{l^+ \,
    l^-} f_2\left( y  \right) \,,
\label{eq:4}
\end{align}
where functions $f_i(y)$ depend on a single kinematical variable $y =
l^2/(l^+ l^-)$. To
determine the functional form of $f_i(y)$, one only needs to calculate
the integrals in Eq.~(\ref{eq:1}) with the constraints
\begin{align}
l^+ = 1 \,, \qquad l^-  = 1 \,.
\end{align}
The full functional dependence can easily be recovered using
Eq.~(\ref{eq:4}). For this reason, we will also denote the auxiliary
integrals as
\begin{align}
  \mi_i(l,p_1, p_2)  \equiv \mi_i ( x ) \,, \qquad x = l^2 \,.
\end{align}
This is an important simplification because the number of independent
scale have been significantly reduced, but not reduced too much such
that there is still nontrivial dependence on the scale.

The auxiliary integrals in Eq.~(\ref{eq:1}) have been calculated to
all order in dimensional regularization parameter in
Ref.~\cite{1105.5171}. 
\begin{align}
  \mi_1 (x) = & \,  \frac{ e^{\e \gamma_E} \Gamma(1 - \e)}{2\, (1-2 \e)
     \Gamma(1-2 \e)} x^{-\e} \,,
\nn\\
  \mi_2 (x) =  &\, - \frac{e^{\e \gamma_E}
    \Gamma(1-\e)}{4 \e \Gamma(1-2\e)} x^{-\e} (1-x)^{-1-\e} {}_2F_1(-\e,-\e;1-\e;x)
 \label{eq:5}
 \end{align}
Results in Eq.~(\ref{eq:5}) can be obtained by parameterizing the phase
space integral in light-cone coordinate and performing the phase space
integral explicitly. However, due to the quadratic constraint imposed
by the delta function $\delta\big( (l-k_1)^2 \big)$, the resulting
integral is not easy to computed. Alternatively, they can be obtained by
dispersion method~\cite{vanNeerven:1985xr}, as along as the discontinuities of the
integrals are well understood. While the dispersion approach is
intrinsically interesting, we prefer not to elaborate it here and
leave it for future work. Instead, we use the method of differential
equation, explained in the next section, for the calculation of such integrals.


\subsection{Solving the auxiliary integrals by differential equation}

Method of differential equation for Feynman integral is originally
proposed for loop integral only. It was then realized that the
same method can also be applied to phase space integral with small
modification~\cite{hep-ph/0207004,hep-ph/0306192}. The reason is that on-shell condition for phase space
integral can be regarded as Feynman propagator for the purpose of IBP
reduction or calculating derivative,
\begin{align}
  \delta_+ ( p^2 )\equiv \frac{1}{(p^2)_c} = \frac{1}{2 \pi i } \left( \frac{1}{p^2 + i0} -
    \frac{1}{p^2 - i0}\right) \,.
\end{align}
Cut propagators are then recovered as delta function after IBP
reduction. The only difference of phase space integral IBP reduction
with conventional loop integral reduction is that integral with zero or
positive power of $p^2$ should be set to zero, because they do not
give rise to discontinuity. The derivation of differential equation
for phase space integral can then be proceeded similar to loop integral.
Taking derivative w.r.t to $x$ can be expressed as
\begin{align}
  \frac{\dd \mi_i(x)}{\dd x} = \frac{1}{2 x} l^\mu \frac{\del
    \mi_i(l,p_1,p_2)}{\del l^\mu}  \,.
\label{eq:6}
\end{align}
The expression on the right-hand side of Eq.~(\ref{eq:6}) can be
reduced  using IBP identities. We obtain
\begin{align}
  \frac{\dd \mi_1(x)}{\dd x}  = & - \frac{\e}{x} \mi_1(x)\,,
\nn\\
  \frac{\dd \mi_2(x)}{\dd x}  = & \frac{ (1-2 \e)}{2\, x \, (1-x)}
  \mi_1(x) + \frac{ (1 +  \e)}{ (1 - x) }  \mi_2(x) \,.
\label{eq:7}
\end{align}
It is a straightforward exercise to solve the system order by order in
$\e$, up to undetermined integration constants. The differential
equation for $\mi_1(x)$ is homogeneous. Its integration constants have
to be fixed by explicit calculation. Using the result given in Eq.~(\ref{eq:5})~\cite{1105.5171},
$\mi_1(x)$ expanded in $\e$ is given by
\begin{align}
  \mi_1 (x) =& \frac{1}{2(1-2 \e)} \bigg[ 
1-\epsilon  H_0(x)+\epsilon ^2 \Big(-\frac{3
  \zeta_2}{2}+H_{0,0}(x)\Big)+\epsilon ^3 \Big(-\frac{7
  \zeta_3}{3}+\frac{3}{2} \zeta_2 H_0(x)-H_{0,0,0}(x)\Big)
\brk
+\epsilon ^4 \Big(-\frac{15 \zeta_4}{16}+\frac{7}{3} \zeta_3 H_0(x)-\frac{3}{2} \zeta_2 H_{0,0}(x)+H_{0,0,0,0}(x)\Big)
\bigg] \,,
\label{eq:8}
\end{align}
where we have only expand through to $\Ord(\e^4)$ for
simplicity. In Eq.~(\ref{eq:8}), $\zeta_n$ is Riemann's zeta value,
$H_{\vec{w}}(x)$ are HPLs, first introduced in physics in
Ref.~\cite{hep-ph/9905237}. In our case, the required HPLs are defined
recursively by
\begin{align}
  H_{0,\vec{w}} (u) = \int^u_0 \! \frac{\dd t}{t} H_{\vec{w}} (t) \,,
  \qquad 
  H_{1,\vec{w}} (u) = \int^u_0 \! \frac{\dd t}{1-t} H_{\vec{w}} (t)
  \,.
\label{eq:30}
\end{align}
If $\vec{w}$ consists of $0$'s only, then it is defined to be $H_{0_n}(u)
= \ln^n u/n!$. As a standard notation, we replace $(k-1)$ 0's followed
by a $1$ with $k$ in the weight vector. For example, $H_{0,0,1} ( u) \equiv  H_{3}(u)$.

Substituting $\mi_1(x)$ into Eq.~(\ref{eq:7}) and solving the
differential equation for $\mi_2(x)$, we obtain
\begin{align}
\mi_2(x) = & \frac{-1}{4\e (1-x)} \bigg[
b_2^{{[0]}}+\epsilon  \Big(b_2^{{[1]}}-H_0(x)+b_2^{{[0]}}
H_1(x)\Big)+\epsilon ^2 \Big(b_2^{{[2]}}+b_2^{{[1]}}
H_1(x)+H_{0,0}(x)-H_{1,0}(x)
\brk
+b_2^{{[0]}} H_{1,1}(x)\Big)+\epsilon ^3 \Big(b_2^{{[3]}}+\frac{3}{2}
\zeta_2 H_0(x)+b_2^{{[2]}} H_1(x)+b_2^{{[1]}}
H_{1,1}(x)-H_{0,0,0}(x)+H_{1,0,0}(x)
\brk
-H_{1,1,0}(x)+b_2^{{[0]}} H_{1,1,1}(x)\Big)+\epsilon ^4
\Big(b_2^{{[4]}}+\frac{7}{3} \zeta_3 H_0(x)+b_2^{{[3]}}
H_1(x)-\frac{3}{2} \zeta_2 H_{0,0}(x)+\frac{3}{2} \zeta_2 H_{1,0}(x)
\brk
+b_2^{{[2]}} H_{1,1}(x)+b_2^{{[1]}}
H_{1,1,1}(x)+H_{0,0,0,0}(x)-H_{1,0,0,0}(x)+H_{1,1,0,0}(x)-H_{1,1,1,0}(x)
\brk
+b_2^{{[0]}} H_{1,1,1,1}(x)\Big)
\bigg] \,,
\label{eq:10}
\end{align}
where $b^{[i]}_2$ are integration constants yet to be specified. An
obvious way to fix these constants are calculating them on the
boundary explicitly, $x=0$ and $x=1$. However, $\mi_2(x)$ is
singular in either of these two limits, as can be seen from the known
result, Eq.~(\ref{eq:5}). Setting naively $x=0$ or $x=1$ in the
integrand doesn't work and one has to perform expansion carefully
along the line of the method of expansion by
region~\cite{hep-ph/9711391}. This however requires further study on
the application of the method of expansion by region to phase space integral,
which is beyond the scope of current work. Instead, we fix these
constants by going to $d=6-2 \e$ dimension and requiring that
$\mi_2(x)$ in $6-2 \e$ dimension vanishes in the limit of $x\to
0$. Using dimensional recurrence relation~\cite{hep-th/9606018}~\footnote{We use the
  implementation of these relations in the package \texttt{LiteRed}.},
$\mi_2(x)$ in $6-2 \e$ dimension is given by
\begin{align}
  \mi_2^{6-2 \e}(x) = -\frac{1}{4 \e} \mi_1(x) -
  \frac{  1 - x }{2
     ( 1 - 2 \e)} \mi_2(x) \,.
\label{eq:11}
\end{align}
The condition $\lim_{x\to 0} \mi_2^{6-2 \e}(x) = 0$ gives
\begin{align}
  \vec{b}_2 = \Big\{
1,0,-\frac{3 \zeta _2}{2},-\frac{7 \zeta _3}{3},-\frac{15 \zeta _4}{16}
\Big\} \,,
\label{eq:12}
\end{align}
where  $b^{[i]}_2$ is the $(i+1)$-th element of the vector
$\vec{b}_2$. We have therefore determined the nontrivial auxiliary
integral $\mi_2(x)$ in almost an algebraic way. In particular, the
integration constants are fixed by regularity condition alone.


\subsection{From auxiliary integral to soft phase space integral}

With the auxiliary integrals at hand, we are ready to solve the soft
phase space integrals in Eq.~(\ref{eq:2}) and (\ref{eq:13}), in  which we are actually
interested. To this end, we need to restore the full dependence of
$\mi_i(l,p_1,p_2)$ on $l$, $l^+$ and $l^-$. Using dimensional analysis and scale symmetry, we
have
\begin{align}
  \mi_1 ( l , p_1 , p_2)  =  & \left( \frac{l^2}{y} \right)^{1 - \e}
  \mi_1 (y) \,,
\nn\\
\mi_2 (l, p_1, p_2) = & \frac{1}{l^+ l^-} \left(\frac{l^2}{y}
\right)^{-\e} \mi_2(y) \,,
\end{align}
where
\begin{align}
  y \equiv \frac{(2\, p_1\mcdot p_2)\,l^2}{(2l^+)\,(2 l^-)} \,,
\end{align}
and $\mi_1(y)$ and $\mi_2(y)$ are given in Eqs.~(\ref{eq:8}) and
(\ref{eq:10}). Under light-cone decomposition, the original integrals
in Eqs.~(\ref{eq:2}) and (\ref{eq:13}) can be written as
\begin{align}
  I_i = & \frac{e^{\e \gamma_E}}{2\,\pi^{d/2-1}} \int\! \dd l^+ \, \dd
  l^- \, \dd^{d-2} l_\perp \, \Theta( l^+ l^- - l^2) \, \delta( 1 -
  l^+ - l^-)  \mi_i(l,p_1,p_2) \,.
\label{eq:20}
\end{align}
Since neither the integrand nor the delta function constrain depends
on $l_\perp$, we can integrate the transverse angular components out explicitly
\begin{align}
  \int\! \dd^{d-2} l_\perp  = \Omega (d - 3) \int\! 
  |l_\perp|^{d-3} \dd |l_\perp| \,,
\end{align}
where
\begin{align}
  \Omega(D) = \frac{2 \pi^{(D+1)/2}}{\Gamma\big((D+1)/2\big)} \,.
\end{align}
Introducing the parameterization
\begin{align}
  l^2 \equiv l^+ l^- - |l_\perp|^2 = u l^+ l^- \,,
\end{align}
$l^+$ and $l^-$ can be easily integrated out in closed form, because
$\mi_i(y)$ are functions of $u$ only,
\begin{align}
  y = \frac{2 \, p_1\mcdot p_2}{(2l^+)\,(2l^-)} u l^+ l^- = u \,.
\end{align}
For $I_1$ the remaining integral in $u$ can be readily carried out order by
order in $\e$. However, additional complexity arises for $I_2$, for
which the explicit integral can be written as
\begin{align}
  I_2 = \frac{e^{\e \gamma_E} \Omega(1 - 2 \e)}{ \pi^{1-\e}}
  \int^{1}_0\! \dd l^+ \, \int^{1}_0 \! \dd u\, \Big( l^+ ( 1 -
  l^+)\Big)^{-1-2 \e} ( 1 - u)^{-\e} u^{-1} \mi_2(u) \,.
\label{eq:9}
\end{align}
We note that the integrand contains an unregulated singularity at
$u=0$. To regulate the integral, one approach is to extract the $\ln u$
dependence of $\mi_2(u)$ to higher order in $\e$ and resum it as $u^{a\e}$,
and then apply the method of sector 
decomposition~\cite{hep-ph/0004013} for the integral.  Alternatively,
the singularity becomes a spurious one in $6 -2 \e$ dimension, and we
can calculate it directly without any regularization procedure.
We can then use dimensional recurrence relation for the soft phase
space integral to obtain the result in
$4 - 2 \e$ dimension. The advantage of this approach is that there is
no power divergence in the integrand in $6-2\e$ dimension, and the
integral can be straightforwardly carried out order by order in
$\e$. Similar strategy has been employed in the direct calculation
of soft phase space integral~\cite{1302.4379}, in the calculation of
multi-box diagrams~\cite{1402.1024}, and in searching for quasi
finite master integral basis~\cite{1411.7392}. In $d=6-2 \e$ dimension,
Eq.~(\ref{eq:9}) is now given by
\begin{align}
    I_2 = \frac{e^{\e \gamma_E} \Omega(3 - 2 \e)}{ \pi^{2-\e}}
  \int^{1}_0\! \dd l^+ \, \int^{1}_0 \! \dd u\, \Big( l^+ ( 1 -
  l^+)\Big)^{1-2 \e} ( 1 - u)^{1-\e} u^{-1} \mi^{6-2 \e}_2(u) \,,
\label{eq:14}
\end{align}
where we have kept the factor $e^{\e \gamma_E}$ unchanged under
dimensional shift by
convention. Eq.~(\ref{eq:14}) still contains the singular factor $1/u$
in the integrand. However, since the function $\mi_2^{6-2 \e}(u)$
vanishes in the limit of $u\to 0$, the integral is finite. We can then
expand the integral order by order in $\e$, and integrate in the
variable $u$ using the definition of HPLs in Eq.~(\ref{eq:30}) and
integration-by-parts relations.  In
this way we obtain results for the original soft phase space integral
in $d=6- 2 \e$ dimension,
\begin{align}
  I_1^{6-2 \e} = & 
\frac{1}{20160} \bigg[
1+\frac{293 \epsilon }{35}+\epsilon ^2 \Big(\frac{508636}{11025}-7
\zeta_2\Big)+\epsilon ^3 \Big(\frac{27711436}{128625}-\frac{293
  \zeta_2}{5}-\frac{62 \zeta_3}{3}\Big)
\brk
+\epsilon ^4 \Big(\frac{113183745616}{121550625}-\frac{508636
  \zeta_2}{1575}-\frac{18166 \zeta_3}{105}-\frac{9
  \zeta_4}{4}\Big)+\epsilon ^5 \Big(\frac{16478717927248}{4254271875}
\brk
-\frac{27711436 \zeta_2}{18375}-\frac{31535432
  \zeta_3}{33075}+\frac{434 \zeta_2 \zeta_3}{3}-\frac{2637
  \zeta_4}{140}-\frac{1022 \zeta_5}{5}\Big)
\brk
+\epsilon ^6
\Big(\frac{21179542653666496}{1340095640625}-\frac{113183745616
  \zeta_2}{17364375}-\frac{1718109032 \zeta_3}{385875}+\frac{18166
  \zeta_2 \zeta_3}{15}
\brk
+\frac{1922 \zeta_3^2}{9}-\frac{127159 \zeta_4}{1225}-\frac{42778 \zeta_5}{25}-\frac{2473 \zeta_6}{16}\Big)
\bigg]
\,,
\nn\\
I_2^{6-2 \e}  =& 
-\frac{1}{96} 
\bigg[
1-\zeta_2+\epsilon  \Big(\frac{31}{3}-\frac{19 \zeta_2}{3}-5
\zeta_3\Big)+\epsilon ^2 \Big(\frac{601}{9}-\frac{328
  \zeta_2}{9}-\frac{95 \zeta_3}{3}+2 \zeta_4\Big)
\brk
+\epsilon ^3 \Big(\frac{9451}{27}-\frac{5308 \zeta_2}{27}-\frac{1511
  \zeta_3}{9}+\frac{161 \zeta_2 \zeta_3}{3}+\frac{38 \zeta_4}{3}-37
\zeta_5\Big)+\epsilon ^4 \Big(\frac{132817}{81}-\frac{78904
  \zeta_2}{81}
\brk
-\frac{22541 \zeta_3}{27}+\frac{3059 \zeta_2 \zeta_3}{9}+\frac{295 \zeta_3^2}{3}+\frac{2039 \zeta_4}{36}-\frac{703 \zeta_5}{3}+\frac{4859 \zeta_6}{48}\Big)
\bigg]
 \,,
\label{eq:15}
\end{align}
where we have kept the expansion through to transcendental
weight $6$. Then we can convert the results back to $d=4-2 \e$ dimension ones
using the following dimension recurrence relations
\begin{align}
  I_1 = & \frac{8 (-3+2 \epsilon ) (-7+4 \epsilon ) (-5+4 \epsilon )
   I^{6-2\e}_1}{-1+\epsilon } \,,
\nn\\
I_2 = &\frac{4 (-3+4 \epsilon ) (-1+4 \epsilon ) \left(-315+1902 \epsilon -4050
   \epsilon ^2+3964 \epsilon ^3-1824 \epsilon ^4+320 \epsilon ^5\right)
   I^{6-2\e}_1}{(-1+\epsilon ) \epsilon ^3 (-1+2 \epsilon )}
\brk
-\frac{4
   (-1+\epsilon ) (-3+4 \epsilon ) (-1+4 \epsilon )
   I_2^{6-2\e}}{\epsilon } \,.
\label{eq:16}
\end{align}
Substituting Eq.~(\ref{eq:15}) into Eq.~(\ref{eq:16}), we obtain
\begin{align}
  I_1 = & \frac{1}{24}
\bigg[
1+\frac{22 \epsilon }{3}+\epsilon ^2 \Big(\frac{340}{9}-7
\zeta_2\Big)+\epsilon ^3 \Big(\frac{4600}{27}-\frac{154
  \zeta_2}{3}-\frac{62 \zeta_3}{3}\Big)+\epsilon ^4
\Big(\frac{58576}{81}-\frac{2380 \zeta_2}{9}
\brk
-\frac{1364 \zeta_3}{9}-\frac{9 \zeta_4}{4}\Big)+\epsilon ^5
\Big(\frac{724192}{243}-\frac{32200 \zeta_2}{27}-\frac{21080
  \zeta_3}{27}+\frac{434 \zeta_2 \zeta_3}{3}-\frac{33
  \zeta_4}{2}-\frac{1022 \zeta_5}{5}\Big)
\brk
+\epsilon ^6 \Big(\frac{8822080}{729}-\frac{410032 \zeta_2}{81}-\frac{285200 \zeta_3}{81}+\frac{9548 \zeta_2 \zeta_3}{9}+\frac{1922 \zeta_3^2}{9}-85 \zeta_4-\frac{22484 \zeta_5}{15}-\frac{2473 \zeta_6}{16}\Big)
\bigg]
\,,
\nn\\
I_2 = & 
-\frac{3}{16 \e^3} 
\bigg[
1-\frac{23}{3} \epsilon ^2 \zeta_2-24 \epsilon ^3 \zeta_3-\frac{11}{12} \epsilon ^4 \zeta_4+\epsilon ^5 \Big(\frac{1624 \zeta_2 \zeta_3}{9}-\frac{3436 \zeta_5}{15}\Big)+\epsilon ^6 \Big(\frac{2512 \zeta_3^2}{9}-\frac{12539 \zeta_6}{144}\Big)
\bigg]
\label{eq:17}
\end{align}
These are the correct results for the NNLO soft phase space
integral. We have seen in this section that the combined use of differential equation,
integral of HPLs, and dimensional recurrence relation leads to an
almost algebraic way for the determination of soft phase space
integral. We will apply this method to the calculation of soft
triple-emission phase space integrals in next section.

\section{Soft triple-emission  phase space integral}
\label{sec:triple}

For soft-virtual corrections at \n3lo, we need to calculate phase
space integrals with three soft partons emitted in the final state. After
reduction by IBP identities, we find a set of $7$ master integrals need to be calculated~\cite{1412.2771},
\begin{align}
  J_1 = & \int\! [dk]  \,,
\nn\\
J_2 = & \int\! [dk]\, \frac{1}{ (k_1+k_3)^2\, \big(2(k_1+k_2)^+\big) } \,,
\nn\\
J_3 = & \int\! [dk] \,\frac{1}{(k_1+k_2+k_3)^2 \, (2 k_2^-) \,   \big(2
  (k_1+k_3)^+\big) } \,,
\nn\\
J_4 = & \int\! [dk] \,\frac{1}{ (k_1+k_3)^2 \,
  (k_1+k_2+k_3)^2\, (2 k_1^+) \, (2 k_2^-)  \, \big(2 (k_2+k_3)^-\big)
} \,,
\nn\\
J_5 = & \int\! [dk] \, \frac{1}{ (k_1+k_2)^2 (k_1+k_3)^2 \, (2 k_2^+) \, (2
  k_3^-) } \,,
\nn\\
J_6 = & \int\! [dk]\, \frac{1}{(k_1+k_2)^2 (k_1+k_3)^2 \, \big(2(k_1+k_2)^+\big) \,
  \big(2 (k_1+k_3)^-\big)} \,,
\nn\\
J_7 = & \int [dk] \,\frac{1}{(k_1+k_2)^2 \, (2 k_1^+) \, (2 k_2^-) }
\,,
\label{eq:18}
\end{align}
where we have used the short-hand notation
\begin{align}
    \int \![dk]  \equiv \frac{e^{3 \e \gamma_E }}{\pi^{3d/2 - 3}} \int\! d^dk_1 \,d^dk_2 \, d^dk_3\delta_+ ( k^2_1) \delta_+
  (k^2_2) \delta_+(k^2_3) \delta( 1 - (k_1+k_2+k_3)\cdot
  (p_1+p_2) )  \,.
\label{eq:19}
\end{align}
The calculation of these integrals follows closely the method given in
Sec.~\ref{sec:diff-equat-appr}.

\subsection{Auxiliary integrals for soft triple-emission phase space integrals}

\label{sec:auxil-integr-triple}

\begin{figure}
  \centering
\includegraphics[width=0.6\textwidth]{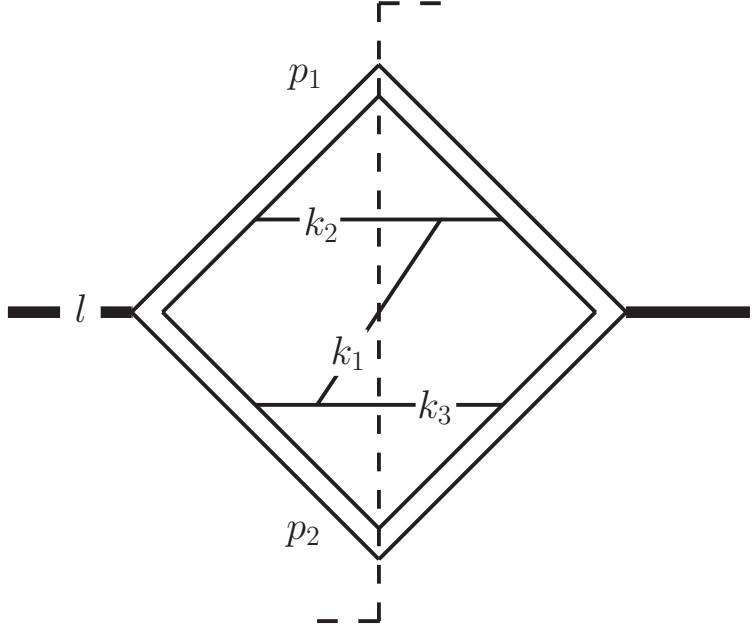}
  \caption{Auxiliary topology for the auxiliary integrals. Dash line is
    phase space cut, and double line
  are eikonal propagators.}
\label{fig:1}
\end{figure}

As in the double-emission case, we introduce internal auxiliary scale
by inserting into Eq.~(\ref{eq:18}) a delta function,
\begin{align}
  1 = \int\! \dd^d l \, \delta^{(d)} \big( l - (k_1 + k_2 + k_3) \big) \,.
\end{align}
To define the auxiliary integrals, it is convenient to introduce the
corresponding auxiliary topology in Fig.~\ref{fig:1}. Auxiliary
integral can be labeled by the power of its individual propagator,
\begin{align}
  \mj(x) = & \;\mathrm{Topo}( j_1, j_2, \ldots, j_9 )
\nn\\
 \equiv &\;\frac{e^{2 \e \gamma_E}}{\pi^{d-2}}  \int\! \dd^d k_1 \,
 \dd^d k_2  \, \dd^d k_3 \, \delta^{(d)}\big( l - (k_1+k_2+k_3) \big)
 \frac{1}{\big[k_1^2\big]^{j_1}_c \, \big[k_2^2\big]^{j_2}_c \, \big[k_3^2\big]^{j_3}_c }
\nn\\
& \; \times \frac{1}{\big[ (k_1+k_2)^2\big]^{j_4} \, \big[(k_1+
  k_3)^2\big]^{j_5} \, \big[2(k_1 + k_2)^+ \big]^{j_6} \, \big[ 2
  (k_1+k_3)^-\big]^{j_7} \, \big[2 k_2^+ \big]^{j_8} \, \big[ 2 k_3^-
  \big]^{j_9}}
\end{align}
We can then write the auxiliary integrals as
\begin{align}
  \mj_1 (x)= & \;\mathrm{Topo}(1,1,1,0,0,0,0,0,0)
\nn\\
= &\; \frac{e^{2 \e \gamma_E}}{\pi^{d-2}} \int\! \dd^d
  k_1 \, \dd^d k_2 \,\delta_+ ( k^2_1) \delta_+
  (k^2_2) \delta_+\big((l-k_1-k_2)^2\big) \,,
\nn\\
  \mj_2 (x)=  & \; \mathrm{Topo}(1,1,1, 0, 1, 1, 0, 0, 0)
\nn\\
= &\; \frac{e^{2 \e \gamma_E}}{\pi^{d-2}} \int\! \dd^d
  k_1 \, \dd^d k_2 \,\frac{\delta_+ ( k^2_1) \delta_+
  (k^2_2) \delta_+\big((l-k_1-k_2)^2\big)}{(l-k_2)^2 \big(2
  (k_1+k_2)^+\big)} \,,
\nn\\
  \mj_3 (x)= &\; \mathrm{Topo}(1,1,1, 0, 0, 0, 1, 1, 0)
\nn\\
 = &\; \frac{e^{2 \e \gamma_E}}{\pi^{d-2}} \int\! \dd^d
  k_1 \, \dd^d k_2 \,\frac{\delta_+ ( k^2_1) \delta_+
  (k^2_2) \delta_+\big((l-k_1-k_2)^2\big)}{ (2 k_2^-) \big(2
  (l-k_2)^+\big)} \,,
\nn\\
  \mj_4 (x)= &\; \mathrm{Topo}(1,1,1, 0, 1, 1, 0, 1, 1)
\nn\\
= & \; \frac{e^{2 \e \gamma_E}}{\pi^{d-2}} \int\! \dd^d
  k_1 \, \dd^d k_2 \,\frac{\delta_+ ( k^2_1) \delta_+
  (k^2_2) \delta_+\big((l-k_1-k_2)^2\big)}{(l-k_2)^2 (2 k_1^+) (2
  k_2^-) \big(2
  (l-k_1)^+\big)} \,,
\nn\\
  \mj_5 (x)= & \; \topo(1, 1, 1, 1, 1, 0, 0, 1, 1)
\nn\\
= &\; \frac{e^{2 \e \gamma_E}}{\pi^{d-2}} \int\! \dd^d
  k_1 \, \dd^d k_2 \,\frac{\delta_+ ( k^2_1) \delta_+
  (k^2_2) \delta_+\big((l-k_1-k_2)^2\big)}{(k_1+k_2)^2 (l-k_2)^2 (2
  k_2^+) \big( 2 (l-k_2)^- \big)} \,,
\nn\\
  \mj_6 (x)= &\; \topo (1, 1, 1, 1, 1, 1, 1, 0, 0)
\nn\\
= &\; \frac{e^{2 \e \gamma_E}}{\pi^{d-2}} \int\! \dd^d
  k_1 \, \dd^d k_2 \,\frac{\delta_+ ( k^2_1) \delta_+
  (k^2_2) \delta_+\big((l-k_1-k_2)^2\big)}{ (k_1+k_2)^2 (l-k_2)^2 \big(2
  (k_1+k_2)^+\big) \big( 2 (l- k_2)^- \big)} \,,
\label{eq:28}
\end{align}
where they correspond to the original integrals $J_{1-6}$. The
calculation of $J_7$ requires additional care and will be addressed
later in Sec.~\ref{sec:fact-integr}.
The calculation of the auxiliary integrals in Eq.~(\ref{eq:28})
follows closely the procedure
illustrated in Sec.~\ref{sec:diff-equat-appr}. We present the results
here and leave the details to appendix~\ref{sec:solv-diffr-equat},
\begin{align}
  \mj_1 (x) = & \frac{x}{4(1 - 2 \e) ( 2 - 9 \e- 9 \e^2) }
\bigg[
1-2 \epsilon  H_0(x)+\epsilon ^2 \Big(-5 \zeta_2+4 H_{0,0}(x)\Big)
+\epsilon ^3 \Big(-\frac{32 \zeta_3}{3}
\brk
+10 \zeta_2 H_0(x)-8 H_{0,0,0}(x)\Big)+\epsilon ^4 \Big(\frac{31 \zeta_4}{4}+\frac{64}{3} \zeta_3 H_0(x)-20 \zeta_2 H_{0,0}(x)+16 H_{0,0,0,0}(x)\Big)
\bigg] \,,
\nn\\
\mj_2 (x) = & \frac{1}{16\e^3} 
\bigg[
-2 \epsilon ^2 \zeta_2+\epsilon ^3 \Big(-14 \zeta_3+4 \zeta_2 H_0(x)\Big)+\epsilon ^4 \Big(-35 \zeta_4+28 \zeta_3 H_0(x)-8 \zeta_2 H_{0,0}(x)\Big)
\bigg] \,,
\nn\\
\mj_3 (x) = & \frac{1}{ 48 ( 1 - 2 \e) \e^2} 
\bigg[
-3 \epsilon  H_1(x)+\epsilon ^2 \Big(6 H_{1,0}(x)-6
H_{1,1}(x)\Big)+\epsilon ^3 \Big(15 \zeta_2 H_1(x)-12 H_{1,0,0}(x)
\brk
+12 H_{1,1,0}(x)-12 H_{1,1,1}(x)\Big)+\epsilon ^4 \Big(32 \zeta_3
H_1(x)-30 \zeta_2 H_{1,0}(x)+30 \zeta_2 H_{1,1}(x)
\brk
+24 H_{1,0,0,0}(x)-24 H_{1,1,0,0}(x)+24 H_{1,1,1,0}(x)-24 H_{1,1,1,1}(x)\Big)
\bigg] \,,
\nn\\
\mj_4 (x) = & \frac{-1}{16\e^3 ( 1 - x)} 
\bigg[
1+\epsilon  \Big(-2 H_0(x)+\frac{H_1(x)}{2}\Big)+\epsilon ^2
\Big(-\frac{9 \zeta_2}{2}-2 H_2(x)+4 H_{0,0}(x)-H_{1,0}(x)
\brk
-2 H_{1,1}(x)\Big)+\epsilon ^3 \Big(-\frac{43 \zeta_3}{6}+9 \zeta_2
H_0(x)-\frac{3}{2} \zeta_2 H_1(x)+2 H_3(x)-4 H_{1,2}(x)+4 H_{2,0}(x)
\brk
-4 H_{2,1}(x)-8 H_{0,0,0}(x)+2 H_{1,0,0}(x)+4 H_{1,1,0}(x)-10
H_{1,1,1}(x)\Big)+\epsilon ^4 \Big(\frac{33 \zeta_4}{2}
\brk
+\frac{43}{3} \zeta_3 H_0(x)+\frac{5}{3} \zeta_3 H_1(x)+12 \zeta_2
H_2(x)-2 H_4(x)-18 \zeta_2 H_{0,0}(x)+3 \zeta_2 H_{1,0}(x)
\brk
+12 \zeta_2 H_{1,1}(x)+4 H_{1,3}(x)-2 H_{2,2}(x)-4 H_{3,0}(x)+4
H_{3,1}(x)-8 H_{1,1,2}(x)+8 H_{1,2,0}(x)
\brk
-8 H_{1,2,1}(x)-8 H_{2,0,0}(x)+8 H_{2,1,0}(x)-8 H_{2,1,1}(x)+16
H_{0,0,0,0}(x)-4 H_{1,0,0,0}(x)
\brk
-8 H_{1,1,0,0}(x)+20 H_{1,1,1,0}(x)-32 H_{1,1,1,1}(x)\Big)
\bigg] \,,
\nn\\
\mj_5 (x) = & \frac{-1}{16\e^3 ( 1 - x) x} 
\bigg[
5+\epsilon  \Big(-10 H_0(x)+7 H_1(x)\Big)+\epsilon ^2 \Big(-25
\zeta_2-6 H_2(x)+20 H_{0,0}(x)
\brk
-14 H_{1,0}(x)+8 H_{1,1}(x)\Big)+\epsilon ^3 \Big(-\frac{148
  \zeta_3}{3}+50 \zeta_2 H_0(x)-35 \zeta_2 H_1(x)+4 H_3(x)
\brk
-12 H_{1,2}(x)+12 H_{2,0}(x)-8 H_{2,1}(x)-40 H_{0,0,0}(x)+28
H_{1,0,0}(x)-16 H_{1,1,0}(x)
\brk
+4 H_{1,1,1}(x)\Big)+\epsilon ^4 \Big(\frac{275
  \zeta_4}{4}+\frac{296}{3} \zeta_3 H_0(x)-\frac{200}{3} \zeta_3
H_1(x)+30 \zeta_2 H_2(x)-100 \zeta_2 H_{0,0}(x)
\brk
+70 \zeta_2 H_{1,0}(x)-40 \zeta_2 H_{1,1}(x)+8 H_{1,3}(x)+8
H_{2,2}(x)-8 H_{3,0}(x)+8 H_{3,1}(x)
\brk
-24 H_{1,1,2}(x)+24 H_{1,2,0}(x)-16 H_{1,2,1}(x)-24 H_{2,0,0}(x)+16
H_{2,1,0}(x)-8 H_{2,1,1}(x)
\brk
+80 H_{0,0,0,0}(x)-56 H_{1,0,0,0}(x)+32 H_{1,1,0,0}(x)-8 H_{1,1,1,0}(x)-16 H_{1,1,1,1}(x)\Big)
\bigg] \,,
\nn\\
\mj_6 (x) = & \frac{-1}{16\e^3 ( 1 - x) x} 
\bigg[
1+\epsilon  \Big(-2 H_0(x)+3 H_1(x)\Big)+\epsilon ^2 \Big(-5 \zeta_2+4
H_{0,0}(x)-6 H_{1,0}(x)
\brk
+9 H_{1,1}(x)\Big)+\epsilon ^3 \Big(-\frac{38 \zeta_3}{3}+10 \zeta_2
H_0(x)-16 \zeta_2 H_1(x)-4 H_3(x)+H_{1,2}(x)-8 H_{0,0,0}(x)
\brk
+12 H_{1,0,0}(x)-18 H_{1,1,0}(x)+27 H_{1,1,1}(x)\Big)+\epsilon ^4
\Big(-\frac{35 \zeta_4}{4}+\frac{76}{3} \zeta_3 H_0(x)-43 \zeta_3
H_1(x)
\brk
-6 \zeta_2 H_2(x)+8 H_4(x)-20 \zeta_2 H_{0,0}(x)+32 \zeta_2
H_{1,0}(x)-50 \zeta_2 H_{1,1}(x)-10 H_{1,3}(x)
\brk
+6 H_{2,2}(x)+8 H_{3,0}(x)-12 H_{3,1}(x)+5 H_{1,1,2}(x)-2
H_{1,2,0}(x)+3 H_{1,2,1}(x)+16 H_{0,0,0,0}(x)
\brk
-24 H_{1,0,0,0}(x)+36 H_{1,1,0,0}(x)-54 H_{1,1,1,0}(x)+81 H_{1,1,1,1}(x)\Big)
\bigg] \,,
\end{align}
where we have given the results through to transcendental weight $4$
only. The corresponding expression through to weight $6$ in electric
form can be requested from the author.

Our next step is to integrate these auxiliary integrals over the
auxiliary scales to obtain soft triple-emission phase space integral.
As was explained in Sec.~\ref{sec:diff-equat-appr}, to further
integrate the auxiliary integrals, we need to go do $d=6-2 \e$
dimension. Using dimension recurrence relations generated
automatically by \texttt{LiteRed}, we obtain the auxiliary
integrals in $6-2\e$ dimension. After restoring the full kinematic
dependence of the auxiliary integrals using scaling symmetry and
dimensional analysis, we can straightforwardly calculate the $6-2\e$
dimension soft phase space integrals using Eq.~(\ref{eq:20}). Through
to weight $6$, they are given by
\begin{align}
  J_1^{6-2\e} = & \frac{1}{319334400}+\frac{69851 \epsilon
  }{1475324928000}+\epsilon ^2
  \Big(\frac{3006159223}{6816001167360000}-\frac{\zeta_2}{19353600}\Big)
\brk
+\epsilon ^3
\Big(\frac{4191505733903}{1259597015728128000}-\frac{69851
  \zeta_2}{89413632000}-\frac{71 \zeta_3}{319334400}\Big)
\brk
+\epsilon ^4
\Big(\frac{3281422567097131711}{145483455316598784000000}-\frac{3006159223
  \zeta_2}{413090979840000}-\frac{4959421
  \zeta_3}{1475324928000}+\frac{13 \zeta_4}{243302400}\Big)
\brk
+\epsilon ^5
\Big(\frac{97057530206488860452951}{672133563562686382080000000}-\frac{4191505733903
  \zeta_2}{76339213074432000}-\frac{213437304833
  \zeta_3}{6816001167360000}
\brk
+\frac{71 \zeta_2 \zeta_3}{19353600}+\frac{908063
  \zeta_4}{1124057088000}-\frac{2591 \zeta_5}{532224000}\Big)+\epsilon
^6
\Big(\frac{2784095300405047890300104143}{3105257063659611085209600000000}
\brk
-\frac{3281422567097131711
  \zeta_2}{8817179110096896000000}-\frac{297596907107113
  \zeta_3}{1259597015728128000}+\frac{4959421 \zeta_2
  \zeta_3}{89413632000}+\frac{5041 \zeta_3^2}{638668800}
\brk
+\frac{39080069899 \zeta_4}{5193143746560000}-\frac{180983941
  \zeta_5}{2458874880000}-\frac{219797 \zeta_6}{40874803200}\Big)\,,
\nn\\
J_2^{6-2 \e} = & J_1^{6-2 \e}\bigg[
4950-2970 \zeta_2+\epsilon  \Big(8850+9837 \zeta_2-20790
\zeta_3\Big)+\epsilon ^2 \Big(17658-2403 \zeta_2+68859 \zeta_3
\brk
-89100 \zeta_4\Big)+\epsilon ^3 \Big(34650-243 \zeta_2-16821
\zeta_3-23760 \zeta_2 \zeta_3+295110 \zeta_4-276210 \zeta_5\Big)
\brk
+\epsilon ^4 \Big(67890-1179 \zeta_2-1701 \zeta_3+78696 \zeta_2
\zeta_3-83160 \zeta_3^2-72090 \zeta_4+914841 \zeta_5
\brk
-953370 \zeta_6\Big)
\bigg]\,,
\nn\\
J_3^{6-2 \e} = & J_1^{6-2 \e}\bigg[
-3190+1980 \zeta_2+\epsilon  \Big(-11196-4743 \zeta_2+15840
\zeta_3\Big)+\epsilon ^2 \Big(-\frac{285267}{8}-\frac{1449 \zeta_2}{2}
\brk
-37944 \zeta_3+76230 \zeta_4\Big)+\epsilon ^3
\Big(-\frac{1715581}{16}-3168 \zeta_2-5796 \zeta_3+11880 \zeta_2
\zeta_3-\frac{365211 \zeta_4}{2}
\brk
+283140 \zeta_5\Big)+\epsilon ^4 \Big(-\frac{10081707}{32}-5778
\zeta_2-25344 \zeta_3-28458 \zeta_2 \zeta_3+47520
\zeta_3^2-\frac{111573 \zeta_4}{4}
\brk
-678249 \zeta_5+1057815 \zeta_6\Big)
\bigg]\,,
\nn\\
J_4^{6-2 \e} = & J_1^{6-2 \e}\bigg[
-415800+415800 \zeta_2-207900 \zeta_3+\epsilon  \Big(-2029410-257490
\zeta_2+3663045 \zeta_3
\brk
-1767150 \zeta_4\Big)+\epsilon ^2 \Big(-7844571-302634 \zeta_2-2037348
\zeta_3-1455300 \zeta_2 \zeta_3+\frac{37739565 \zeta_4}{2}
\brk
-6444900 \zeta_5\Big)+\epsilon ^3 \Big(-27012717-1367226
\zeta_2-1888776 \zeta_3+6722415 \zeta_2 \zeta_3-6029100 \zeta_3^2
\brk
-9721503 \zeta_4+65945295 \zeta_5-\frac{60516225 \zeta_6}{2}\Big)
\bigg]\,,
\nn\\
J_5^{6-2 \e} = & J_1^{6-2 \e}\bigg[
166320-83160 \zeta_2+\epsilon  \Big(595548+34866 \zeta_2-498960
\zeta_3\Big)+\epsilon ^2 \Big(1781496-32292 \zeta_2
\brk
+209196 \zeta_3-1746360 \zeta_4\Big)+\epsilon ^3 \Big(4791420-23814 \zeta_2-193752 \zeta_3+732186 \zeta_4-4989600 \zeta_5\Big)
\bigg]\,,
\nn\\
J_6^{6-2 \e} = & J_1^{6-2 \e}\bigg[
415800-249480 \zeta_2+\epsilon  \Big(1488870+437238 \zeta_2-1829520
\zeta_3\Big)+\epsilon ^2 \Big(4453740
\brk
+428940 \zeta_2+3095532 \zeta_3-8191260 \zeta_4\Big)+\epsilon ^3
\Big(11978550+1109358 \zeta_2+2970288 \zeta_3
\brk
-2162160 \zeta_2 \zeta_3+13413501 \zeta_4-26611200 \zeta_5\Big)
\bigg] \,.
\end{align}
Finally, using dimension recurrence relations for soft phase space
integral generated by \texttt{LiteRed}, we obtain the results in $4-2 \e$ dimension,
\begin{align}
  J_1 =& \frac{1}{960}+\frac{137 \epsilon }{9600}+\epsilon ^2
  \Big(\frac{12019}{96000}-\frac{11 \zeta_2}{640}\Big)+\epsilon ^3
  \Big(\frac{874853}{960000}-\frac{1507 \zeta_2}{6400}-\frac{71
    \zeta_3}{960}\Big)+\epsilon ^4
  \Big(\frac{58067611}{9600000}
\brk
-\frac{132209
    \zeta_2}{64000}-\frac{9727 \zeta_3}{9600}+\frac{91
    \zeta_4}{5120}\Big)+\epsilon ^5
  \Big(\frac{3673451957}{96000000}-\frac{9623383
    \zeta_2}{640000}-\frac{853349 \zeta_3}{96000}+\frac{781 \zeta_2
    \zeta_3}{640}
\brk
+\frac{12467 \zeta_4}{51200}-\frac{2591
    \zeta_5}{1600}\Big)+\epsilon ^6
  \Big(\frac{226576032859}{960000000}-\frac{638743721
    \zeta_2}{6400000}-\frac{62114563 \zeta_3}{960000}+\frac{106997
    \zeta_2 \zeta_3}{6400}
\brk
+\frac{5041 \zeta_3^2}{1920}+\frac{1093729
    \zeta_4}{512000}-\frac{354967 \zeta_5}{16000}-\frac{219797
    \zeta_6}{122880}\Big)
\,,
\nn\\
J_2 = & -J_1 \frac{30}{\e} \bigg[
\zeta_2+\epsilon  \Big(-\frac{47 \zeta_2}{10}+7 \zeta_3\Big)+\epsilon
^2 \Big(\frac{36 \zeta_2}{5}-\frac{329 \zeta_3}{10}+30
\zeta_4\Big)+\epsilon ^3 \Big(-\frac{18 \zeta_2}{5}+\frac{252
  \zeta_3}{5}
\brk
+8 \zeta_2 \zeta_3-141 \zeta_4+93 \zeta_5\Big)
\bigg]\,, 
\nn\\
J_3 = & - J_1 \frac{30}{\e} \bigg[
\zeta_2+\epsilon  \Big(-\frac{57 \zeta_2}{10}+8 \zeta_3\Big)+\epsilon
^2 \Big(\frac{99 \zeta_2}{10}-\frac{228 \zeta_3}{5}+\frac{77
  \zeta_4}{2}\Big)+\epsilon ^3 \Big(-\frac{27 \zeta_2}{5}+\frac{396
  \zeta_3}{5}
\brk
+6 \zeta_2 \zeta_3-\frac{4389 \zeta_4}{20}+143 \zeta_5\Big)+\epsilon ^4 \Big(-\frac{216 \zeta_3}{5}-\frac{171 \zeta_2 \zeta_3}{5}+24 \zeta_3^2+\frac{7623 \zeta_4}{20}-\frac{8151 \zeta_5}{10}+\frac{2137 \zeta_6}{4}\Big)
\bigg]\,, 
\nn\\
J_4 = & -  J_1  \frac{20}{\e^5}\bigg[
1-\frac{107 \epsilon }{10}+\epsilon ^2 \Big(\frac{87}{5}-\frac{3
  \zeta_2}{4}\Big)+\epsilon ^3 \Big(\frac{999}{5}+\frac{321
  \zeta_2}{40}-\frac{3 \zeta_3}{4}\Big)+\epsilon ^4
\Big(-\frac{6777}{5}-\frac{261 \zeta_2}{20}
\brk
+\frac{321 \zeta_3}{40}+\frac{63 \zeta_4}{4}\Big)+\epsilon ^5
\Big(\frac{24867}{5}-\frac{2997 \zeta_2}{20}-\frac{261 \zeta_3}{20}+12
\zeta_2 \zeta_3-\frac{6741 \zeta_4}{40}+\frac{393 \zeta_5}{4}\Big)
\brk
+\epsilon ^6 \Big(-15309+\frac{20331 \zeta_2}{20}-\frac{2997
  \zeta_3}{20}-\frac{642 \zeta_2 \zeta_3}{5}+60 \zeta_3^2+\frac{5481
  \zeta_4}{20}-\frac{42051 \zeta_5}{40}+\frac{2063 \zeta_6}{4}\Big) 
\bigg]\,,
\nn\\
J_5 = & - J_1 \frac{90}{\e^5} \bigg[
1-\frac{137 \epsilon }{10}+\epsilon ^2 \Big(\frac{135}{2}-2
\zeta_2\Big)+\epsilon ^3 \Big(-153+\frac{137 \zeta_2}{5}-12
\zeta_3\Big)+\epsilon ^4 \Big(162-135 \zeta_2
\brk
+\frac{822 \zeta_3}{5}-42 \zeta_4\Big)+\epsilon ^5
\Big(-\frac{324}{5}+306 \zeta_2-810 \zeta_3+\frac{2877 \zeta_4}{5}-120
\zeta_5\Big)+\epsilon ^6 \Big(-324 \zeta_2
\brk
+1836 \zeta_3-2835 \zeta_4+1644 \zeta_5-310 \zeta_6\Big)
\bigg]\,, 
\nn\\
J_6 = &  - J_1 \frac{15}{\e^5} \bigg[
1-\frac{137 \epsilon }{10}+\epsilon ^2 \Big(\frac{135}{2}-4
\zeta_2\Big)+\epsilon ^3 \Big(-153+\frac{274 \zeta_2}{5}-40
\zeta_3\Big)+\epsilon ^4 \Big(162-270 \zeta_2
\brk
+548 \zeta_3-222 \zeta_4\Big)+\epsilon ^5 \Big(-\frac{324}{5}+612
\zeta_2-2700 \zeta_3-56 \zeta_2 \zeta_3+\frac{15207 \zeta_4}{5}-864
\zeta_5\Big)
\brk
+\epsilon ^6 \Big(-648 \zeta_2+6120 \zeta_3+\frac{3836 \zeta_2 \zeta_3}{5}-196 \zeta_3^2-14985 \zeta_4+\frac{59184 \zeta_5}{5}-3460 \zeta_6\Big)
\bigg]\,.
\label{eq:21}
\end{align}
It is interesting to note that the rational part of $J_5$ and $J_6$
equals up to an overall normalization. 
\subsection{The factorizable integral}
\label{sec:fact-integr}

We still have the last integral $J_7$ to calculate. While it is
straightforward to solve the corresponding auxiliary integral using differential equation, it
is however difficult to fix the integration constants. In particular, the
regularity condition does not fix the integration constants. To
calculate this integral, we observe that the set of momentum
$\{k_1,k_2\}$ and $\{ k_3\}$ are entangled together only by the delta function,
\begin{align}
  \delta^{(d)} \big( 1 - (k_1 + k_2 + k_3) \mcdot (p_1+p_2) \big) \,.
\label{eq:22}
\end{align}
It is well-known in resummation study that such entanglement can be
factorized into a
double-emission integral and a single-emission integral via
Laplace/Fourier transformation. Even more straightforwardly, we can consider a one parameter integral
\begin{align}
  J_7(\omega) = & \frac{e^{3 \e \gamma_E }}{\pi^{3d/2 - 3}} \int\! d^dk_1 \,d^dk_2 \, d^dk_3\,
  \frac{\delta_+ ( k^2_1) \delta_+
  (k^2_2) \delta_+(k^2_3) \delta( \omega - (k_1+k_2+k_3)\cdot
  (p_1+p_2) ) }{(k_1+k_2)^2 \, (2 k_1^+) \, (2 k_2^-) } \,.
\label{eq:23}
\end{align}
The original integral is simply $J_7(1)$. To factorize the integral,
we rewrite it as
\begin{align}
J_7(\omega)  = \int\! \dd \omega_1 \, J_{7,12}(\omega_1) \int\! \dd
\omega_2 \, J_{7,3} ( \omega_2) \, \delta ( \omega - \omega_1 -
\omega_2) \,,
\label{eq:24}
\end{align}
where
\begin{align}
  J_{7,12} (\omega)  = &\frac{e^{2 \e \gamma_E }}{\pi^{d - 2}} \int\! d^dk_1 \,d^dk_2 \,  
 \frac{\delta_+( k^2_1) \delta_+
  (k^2_2)  \delta( \omega - (k_1+k_2)\cdot
  (p_1+p_2) ) }{(k_1+k_2)^2 \, (2 k_1^+) \, (2 k_2^-) } \,,
\nn\\
  J_{7,3} (\omega) = &\frac{e^{ \e \gamma_E }}{\pi^{d/2 - 1}} \int\! d^dk_3  \,  
 \delta_+( k^2_3)  \delta( \omega - k_3\cdot
  (p_1+p_2) )
\end{align}
We note that by dimensional analysis, $J_{7,12}(\omega)$ is
proportional to $I_2$ in
Eq.~(\ref{eq:17}). In $d= 4 - 2 \e$ dimension, we have
\begin{align}
   J_{7,12} (\omega) = \omega^{-1-4\e}  I_2 \,,
\label{eq:26}
\end{align}
and $J_{7,3}(\omega)$ is the simple soft single-emission phase space
integral,
\begin{align}
  J_{7,3}(\omega) =\frac{\omega^{1-2 \e} }{2 (1-2 \e)} \bigg[
1-\frac{3}{2} \epsilon ^2 \zeta_2-\frac{7}{3} \epsilon ^3 \zeta_3-\frac{15}{16} \epsilon ^4 \zeta_4+\epsilon ^5 \Big(\frac{7 \zeta_2 \zeta_3}{2}-\frac{31 \zeta_5}{5}\Big)+\epsilon ^6 \Big(\frac{49 \zeta_3^2}{18}-\frac{399 \zeta_6}{128}\Big)
\bigg] \,.
\label{eq:25}
\end{align}
Substituting Eqs.~(\ref{eq:26}) and (\ref{eq:25}) into
Eq.~(\ref{eq:24}), we obtain
\begin{align}
J_7  =  & J_1 \frac{45}{2 \e^4}  \bigg[
1-\frac{77 \epsilon }{10}+\epsilon ^2 \Big(\frac{213}{10}-\frac{2
  \zeta_2}{3}\Big)+\epsilon ^3 \Big(-\frac{126}{5}+\frac{77
  \zeta_2}{15}-\frac{10 \zeta_3}{3}\Big)+\epsilon ^4
\Big(\frac{54}{5}-\frac{71 \zeta_2}{5}
\brk
+\frac{77 \zeta_3}{3}-\frac{31 \zeta_4}{3}\Big)+\epsilon ^5
\Big(\frac{84 \zeta_2}{5}-71 \zeta_3-\frac{4 \zeta_2
  \zeta_3}{3}+\frac{2387 \zeta_4}{30}-\frac{74
  \zeta_5}{3}\Big)+\epsilon ^6 \Big(-\frac{36 \zeta_2}{5}+84 \zeta_3
\brk
+\frac{154 \zeta_2 \zeta_3}{15}-\frac{10 \zeta_3^2}{3}-\frac{2201 \zeta_4}{10}+\frac{2849 \zeta_5}{15}-\frac{1111 \zeta_6}{18}\Big)
\bigg]
\label{eq:27}
\end{align}

Eqs.~(\ref{eq:21}) and (\ref{eq:27}) are the main results of this work. Similar integrals
have also been computed in Ref.~\cite{1302.4379} using completely different
method and our results agree perfectly with theirs
after taking into account the difference in the normalization.

\section{Conclusion and outlook}
\label{sec:conclusion-outlook}

In this work we have presented the calculation of soft triple-emission
phase space integrals, which are relevant for soft-virtual corrections
to Higgs production at \n3lo~\cite{1403.4616,1412.2771}. The major
techniques used in this calculation are differential equation for
auxiliary integrals, and integration of harmonic polylogarithms. The essential idea is to introduce \emph{internal} scales into
single scale integrals. Since integral in the eikonal limit enjoys
the property of scale invariance, the \emph{internal} scales should be
introduced in a way that doesn't spoil the scale invariance. In this
way the resulting auxiliary integrals obey strong constraints from
scale invariance, and therefore simplify significantly the solution of
differential equations. A particular nice feature of the system of
differential equation is that, all the nontrivial integration
constants can be determined by regularity condition in $6-2 \e$
dimension. Furthermore, the solution of differential equations, as
well as the subsequent integration of auxiliary scales, only involves
the well-known generalization of logarithms and
polylogarithms, the harmonic polylogarithms. Thanks to all these
techniques, we are able to calculate the soft phase space integrals in an
almost algebraic way. The meaning of such a calculation is two
folds. First, it provides an alternative method to the one used in the
calculation of Ref.~\cite{1302.4379}. Second, it provides the first
independent confirmation of the integrals calculated in Ref.~\cite{1302.4379}.

Recently, significantly progresses have been made towards the
calculation of \n3lo QCD corrections to the Higgs boson production via
gluon
fusion~\cite{hep-ph/9708255,hep-ph/0512058,hep-ph/0512060,PHLTA.B93.429,hep-ph/9302208,hep-ph/9701390,hep-ph/0411261,0902.3519,1001.2887,1004.3653,1010.4478,1211.6559,1302.4379,1309.4391,1309.4393,1311.1425,1312.1296,1403.4616,1404.5839,1407.4049,1411.3586,1411.3587,1411.3584}. We
hope that the method and results presented in this work will be useful
in the pursuit of ultimate precision for Higgs physics.

\vskip0.5cm
\noindent {\large\bf Acknowledgments}
\vskip0.3cm

We are grateful to Ye~Li, Andreas~von~Manteuffel, and
Robert~Schabinger for collaboration on related topic. We would like to
thank Robert~Schabinger for useful discussion. The figure was
generated using \texttt{Jaxodraw}~\cite{Binosi:2003yf}, based on \texttt{AxoDraw}~\cite{Vermaseren:1994je}. This work was
supported by the US Department of Energy under contract DE-AC02-76SF00515.

\appendix

\section{Solving diffrential equations for auxiliary integral}
\label{sec:solv-diffr-equat}

In this appendix we solve the auxiliary integrals in Eq.~(\ref{eq:28})
using the method of differential equation. It turns out that the
system of auxiliary integrals in Eq.~(\ref{eq:28}) is not
closed under the derivative of $x$. Additional auxiliary integrals are
needed. A complete master integral basis is given by
\begin{align}
  g_1(x) = & \; -\frac{(2 \e-1) (3 \e-2) (3 \e-1) \mj_1(x)}{x} \,,
\nn\\
g_2(x) = & \; \e^3 \mj_2(x) \,,
\nn\\
g_3(x) = &\; -\e^2 (2 \e-1) \mj_3(x) \,,
\nn\\
g_4(x) = & \; -\e^3 (x-1) \mj_4(x) \,,
\nn\\
g_5(x) = & \; -\e^3 (x-1) x \mj_5(x) \,,
\nn\\
g_6 (x) = & \; -\e^3 (x-1) x \mj_6(x)\,,
\nn\\
g_7 (x) = &\; \e^3 \mj_2(x)-2 \e^3 \topo(1,1,1,0,1,1,0,0,1)-\frac{(2
  \e-1) (3 \e-2) (3 \e-1) \mj_1(x)}{4 x} \,,
\nn\\
g_8(x) = & \; \e^3 \topo(1,1,1,0,0,0,0,1,1) \,,
\nn\\
g_9(x) = &\; \e^3 \topo(1,2,1,0,0,1,1,0,0) \,,
\nn\\
g_{10}(x) = &\; \e^3 \topo(1,1,1,0,0,1,1,0,0) \,,
\nn\\
g_{11}(x) = & \; -\e^2 (x-1) \topo(2,1,1,0,0,1,1,0,0) \,,
\nn\\
g_{12}(x) = & \; \e^3 x \topo(1,1,1,1,1,1,1,-1,0) \,,
\nn\\
g_{13}(x) = & \; -\frac{6 \e^3 \topo(1,1,1,0,0,0,0,1,1)}{x-1}+\e^3 x
\topo(1,1,1,1,1,0,-1,1,1)
\brk
+\frac{3 (2 \e-1) (3 \e-2) (3 \e-1)
  \mj_1(x)}{4 (x-1) x} \,
\end{align}
Note that we have chosen a basis such that the system of differential
equation has \emph{optimal} form~\cite{1304.1806,1412.2296}. The advantage of the optimal basis is
that the resulting differential equation can be solved almost
trivially, up to some integration constants, as demonstrated in many
non-trivial examples~\cite{1306.2799,1306.3504,1306.6344,1307.4083,1312.2588,1401.2979,1404.2922,1404.4853,1404.5839,1407.4049,1408.3107,1408.5134,1409.0023,1410.2804,1411.0911,1411.3586,1411.3587}. Specifically, in $d=4-2 \e$ dimension, the differential equation
for the optimal basis can be written as
\begin{align}
  \frac{\dd \vec{g}(x)}{\dd x} = \e \left( \frac{1}{x} \boldsymbol{A}
    + \frac{1}{1-x} \boldsymbol{B}  \right) \vec{g}(x) \, .
\label{eq:29}
\end{align}
The important feature of optimal basis is that the dependence on $\e$
can be factored out, and $\boldsymbol{A}$ and $\boldsymbol{B}$ are constant matrices 
in field of rational numbers. Since the $\e$ dependence on the
right-hand  side of Eq.~(\ref{eq:29}) has been factored out, it is clear that
the derivative of $\vec{g}(x)$ at $\Ord(\e^{i+1})$ depends only on the
$\Ord(\e^{i})$ terms of $\vec{g}(x)$. Therefore, one can simply solve
Eq.~(\ref{eq:29}) in a bottom-up way by integrating the right-hand
side of Eq.~(\ref{eq:29}). In our case, the matrices $\boldsymbol{A}$
and $\boldsymbol{B}$ are given by
\begin{align}
  \boldsymbol{A} = \left(
\begin{array}{ccccccccccccc}
 -2 & 0 & 0 & 0 & 0 & 0 & 0 & 0 & 0 & 0 & 0 & 0 & 0 \\
 0 & -2 & 0 & 0 & 0 & 0 & 0 & 0 & 0 & 0 & 0 & 0 & 0 \\
 0 & 0 & 0 & 0 & 0 & 0 & 0 & 0 & 0 & 0 & 0 & 0 & 0 \\
 \frac{1}{4} & -\frac{1}{2} & -3 & 0 & 0 & 0 & \frac{1}{2} & 0 & 0 & 0 & 0 &
   0 & 0 \\
 \frac{3}{2} & 0 & 0 & 0 & 0 & 0 & 0 & 6 & 0 & 0 & 0 & 0 & -1 \\
 \frac{4}{5} & -2 & 0 & 0 & 0 & 0 & 0 & 0 & \frac{14}{5} & -\frac{6}{5} &
   \frac{4}{5} & -1 & 0 \\
 -\frac{1}{2} & 0 & 6 & 0 & 0 & 0 & -1 & 0 & 0 & 0 & 0 & 0 & 0 \\
 0 & 0 & 0 & 0 & 0 & 0 & 0 & 0 & 0 & 0 & 0 & 0 & 0 \\
 -\frac{3}{10} & 0 & 0 & 0 & 0 & 0 & 0 & 0 & -\frac{14}{5} & \frac{6}{5} &
   -\frac{4}{5} & 0 & 0 \\
 0 & 0 & 0 & 0 & 0 & 0 & 0 & 0 & 0 & 0 & 0 & 0 & 0 \\
 \frac{21}{20} & 0 & 0 & 0 & 0 & 0 & 0 & 0 & \frac{14}{5} & -\frac{6}{5} &
   \frac{4}{5} & 0 & 0 \\
 -\frac{3}{10} & 0 & 0 & 0 & 0 & 2 & 0 & 0 & -\frac{14}{5} & \frac{6}{5} &
   -\frac{4}{5} & -3 & 0 \\
 \frac{3}{2} & 0 & 0 & 0 & 2 & 0 & 0 & 18 & 0 & 0 & 0 & 0 & -3 \\
\end{array}
\right)
\end{align}
and
\begin{align}
  \boldsymbol{B} = \left(
\begin{array}{ccccccccccccc}
 0 & 0 & 0 & 0 & 0 & 0 & 0 & 0 & 0 & 0 & 0 & 0 & 0 \\
 0 & 0 & 0 & 0 & 0 & 0 & 0 & 0 & 0 & 0 & 0 & 0 & 0 \\
 -\frac{1}{4} & 0 & 2 & 0 & 0 & 0 & 0 & 0 & 0 & 0 & 0 & 0 & 0 \\
 \frac{3}{8} & 0 & -3 & 2 & 0 & 0 & 0 & 0 & 0 & 0 & 0 & 0 & 0 \\
 \frac{3}{4} & 0 & 0 & 0 & 2 & 0 & 0 & 6 & 0 & 0 & 0 & 0 & 0 \\
 0 & 0 & 0 & 0 & 0 & 2 & 0 & 0 & 0 & 0 & 1 & 0 & 0 \\
 0 & 2 & 6 & 0 & 0 & 0 & -1 & 0 & 0 & 0 & 0 & 0 & 0 \\
 \frac{1}{4} & 0 & 0 & 0 & 0 & 0 & 0 & 2 & 0 & 0 & 0 & 0 & 0 \\
 -\frac{3}{10} & 0 & 0 & 0 & 0 & 0 & 0 & 0 & -\frac{9}{5} & \frac{6}{5} &
   -\frac{4}{5} & 0 & 0 \\
 -\frac{1}{5} & 0 & 0 & 0 & 0 & 0 & 0 & 0 & -\frac{6}{5} & \frac{4}{5} &
   -\frac{1}{5} & 0 & 0 \\
 0 & 0 & 0 & 0 & 0 & 0 & 0 & 0 & 0 & 0 & 3 & 0 & 0 \\
 -\frac{3}{10} & 0 & 0 & 0 & 0 & 2 & 0 & 0 & -\frac{4}{5} & \frac{6}{5} &
   -\frac{4}{5} & -1 & 0 \\
 \frac{3}{2} & 0 & 0 & 0 & 2 & 0 & 0 & 18 & 0 & 0 & 0 & 0 & -1 \\
\end{array}
\right) \,.
\end{align}

While integrating the differential equation is trivial with the
optimal basis, determining the integration constants is not. In
particular, the master integrals at phase space point
$x=0$ or $x=1$ are in general singular in $4-2 \e$ dimension, therefore method of asymptotic
expansion is needed. However, for phase space integral, we are not
aware of simple procedure for asymptotic expansion. The solution of
this difficulty is by going to $d=6-2 \e$, where the degree of
singularity is lessen. In particular, we find that integration
constants for all the master integrals, except the ones with trivial
$x$ dependence, $\mj_1(x)$ and $\mj_2(x)$, can all be uniquely determined by the following regularity conditions,
\begin{itemize}
\item There are no power law divergences for the master integrals at
  $x=0$ nor $x=1$.
\item If the master integrals in the $x\to 0$ limit are free of logarithmic
  singularity, then they vanish in that limit. 
\end{itemize}
These conditions can be inferred by studying the asymptotic
expansion of corresponding auxiliary integrals, treating the cut
propagators as normal propagators. For the integration constants of
$\mj_1(x)$ and $\mj_2(x)$, we calculate directly using the dispersion
method, to which we wish to come back in the future.


\end{document}